\pdfoutput=1
 
\documentclass[sigconf, nonacm]{acmart}

\AtBeginDocument{%
  }

\usepackage{times}
\usepackage{latexsym}

\usepackage{multirow}
\usepackage{amsmath}
\usepackage{relsize}
\usepackage{commath}
\usepackage{graphicx}
\usepackage{lingmacros}
\usepackage{enumitem}

\newcommand{\remove}[1]{}

\usepackage[T1]{fontenc}

\usepackage[utf8]{inputenc}

\usepackage{microtype}

\begin{document}

\title[QueryBuilder]{QueryBuilder: Human-in-the-Loop Query Development\\for Information Retrieval}

\author{
Hemanth Kandula$^{1}$, \hspace{0.25cm} Damianos Karakos$^{1}$, \hspace{0.25cm} Haoling Qiu$^{1}$, \hspace{0.25cm} Benjamin Rozonoyer$^{2}$, \\
\ Ian Soboroff$^{3}$, \hspace{0.25cm} Lee Tarlin$^{1}$, \hspace{0.25cm} Bonan Min$^{4}$ \textsuperscript{\dag}\\
Raytheon BBN Technologies$^{1}$ \hspace{0.25cm} University of Massachusetts, Amherst$^{2}$ \\
National Institute of Standards and Technology$^{3}$ \hspace{0.25cm} Amazon AWS AI Labs$^{4}$ \\
{\tt \normalsize \{hemanth.kandula, damianos.karakos, haoling.qiu, lee.tarlin\}@rtx.com} \\
{\tt \normalsize ian.soboroff@nist.gov, brozonoyer@iesl.cs.umass.edu, bonanmin@amazon.com}
}

\renewcommand{\shortauthors}{} 

\begin{abstract}
Frequently, users of an Information Retrieval (IR) system start with an overarching information need (a.k.a., an analytic task) and proceed to define finer-grained queries covering various important aspects (i.e., sub-topics) of that analytic task.
We present a novel, interactive system called {\em QueryBuilder}, which allows a novice, English-speaking user to create queries with a small amount of effort, through efficient exploration of an English development corpus in order to rapidly develop cross-lingual information retrieval queries corresponding to the user's information needs.
QueryBuilder performs near real-time retrieval of documents based on user-entered search terms; the user looks through the retrieved documents and marks sentences as relevant to the information needed. The marked sentences are used by the system as additional information in query formation and refinement: query terms (and, optionally, event features, which capture event 'triggers' (indicator terms) and agent/patient roles) are appropriately weighted, and a neural-based system, which better captures textual meaning, retrieves other relevant content. The process of retrieval and marking is repeated as many times as desired, giving rise to increasingly refined queries in each iteration. The final product is a fine-grained query used in Cross-Lingual Information Retrieval (CLIR). Our experiments using analytic tasks and requests from the IARPA BETTER IR datasets show that with a small amount of effort (at most 10 minutes per sub-topic), novice users can form {\em useful} fine-grained queries including in languages they don't understand. QueryBuilder also provides beneficial capabilities to the traditional corpus exploration and query formation process. A demonstration video is released at \url{https://vimeo.com/734795835}

\end{abstract}




\maketitle
\footnotetext{\textsuperscript{\large\dag} Work done prior to joining AWS AI Labs.}

\section{Introduction}
\label{sec:introduction}
In information retrieval (IR), the user has an information need, which, to be useable by an automated IR system, has to be encoded as a query that the system can process. In the context of the IARPA BETTER IR datasets \cite{soboroff-etal-2023-better}, English queries are created by NIST (as described in a later section) and consist of a narrative (a natural language passage) as well as {\em example} snippets\footnote{This is the ``query-by-example'' scenario.} written in a language that the user can understand, e.g., English. The job of the IR system is to discover what is salient about these pieces of information and use it to search a large foreign document collection to find relevant documents.

Converting the information need into a query (``query development'') is often time-consuming and requires expert knowledge. Frequently, a user starts off with only a vague idea of the information need. We call that the ``overarching analytic task'', which is a general topic such as ``{\bf North Korean ballistic missile and nuclear testing, (2016-2017)}''. However, this task description is not detailed enough to cover all sub-topics of interest to the user. For each sub-topic, the IR system will need a more fine-grained query, containing a more precise description of the information need. For example, consider the following fine-grained queries
which were developed by an expert annotator for the above analytic task: {\em (a) Identify dates and locations of the missile and nuclear tests. (b) Identify capabilities exhibited in missile tests, including range and fuel type. (c) What were the successes and failures of the missile tests?}

\begin{figure}
  \centering
  \includegraphics[scale=0.20]{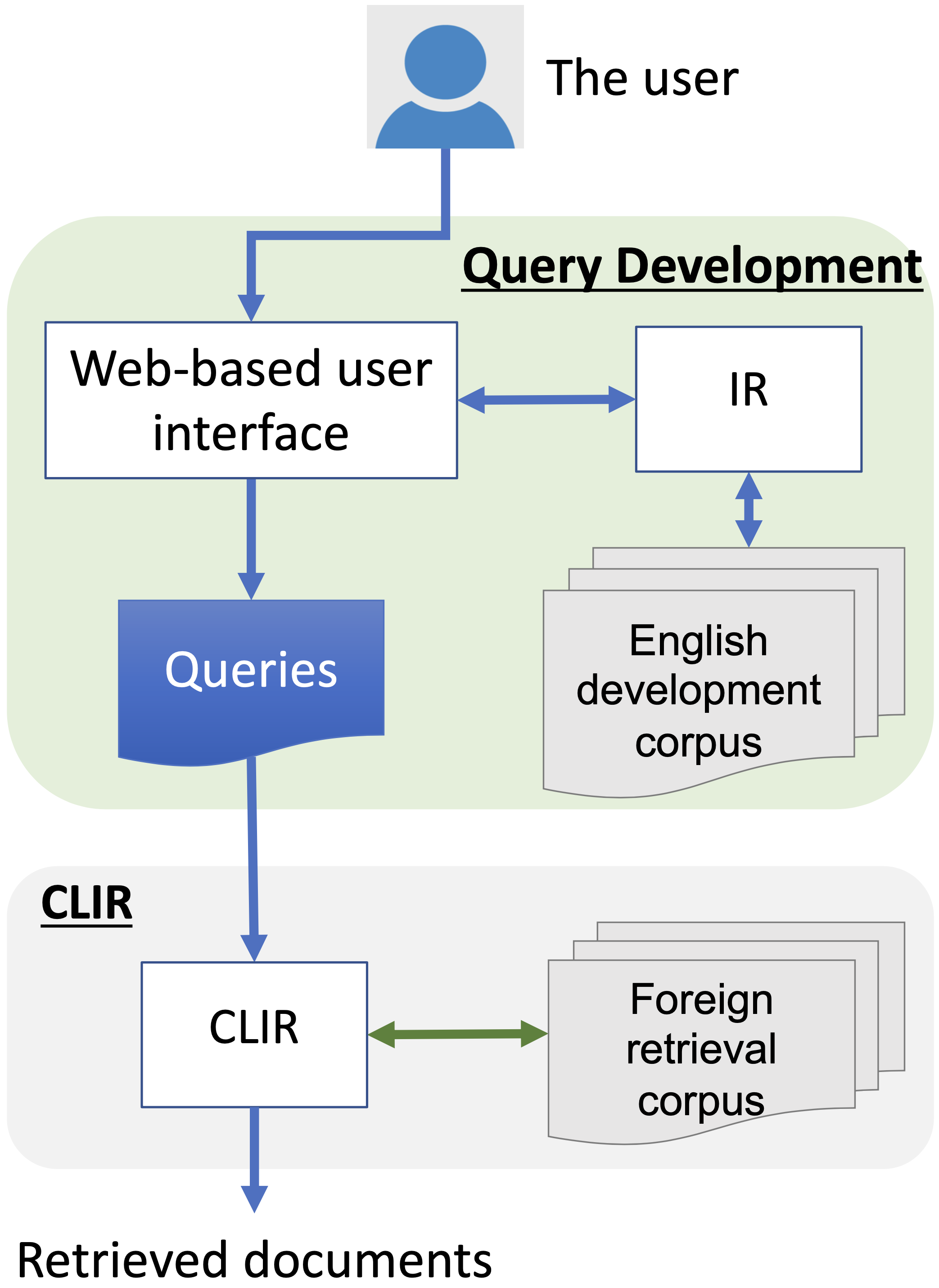}
  \caption{An illustration of the human-in-the-loop query development process and its relation to CLIR.}
  \label{fig:hitl_overall}
  \vspace{-0.15in}
\end{figure}

To develop the sub-topics, analysts/annotators typically have to read through a large set of documents that seem relevant to the overarching task. Such an activity can help the user form a ``query topic landscape'' which can be further refined based on what is ``interesting''. {\em How to efficiently explore the large set of documents and how to refine the information and form the final query is an important problem and is the topic of this paper.} Specifically, we present an interactive system, {\em QueryBuilder}, which allows a novice user, who is not a practitioner of IR, to create queries efficiently, given an overarching task. The key idea is to utilize an efficient IR system for English corpus exploration using a simple list of words as the initial query in the retrieval, and, based on that exploration, find {\it example} snippets~\footnote{A snippet refers to a sentence in this paper. It is straightforward to extend it to other forms such as a clause, a paragraph, or a document.} that are relevant to the topic of interest for subsequent query refinement. The exploration is done iteratively, where each iteration proceeds with an increasingly refined version of the query, based on relevance feedback by the user.
Subsequently, the queries are sent to a Cross-Lingual Information Retrieval (CLIR) system to perform retrieval on a foreign-document collection\footnote{We use a {\em foreign} collection for the final evaluation, to be consistent with the goals of the IARPA BETTER program.}. Figure~\ref{fig:hitl_overall} illustrates the human-in-the-loop (HITL) setting for query development, with key resources and components highlighted. We separate the HITL interactions (the {\it query development} phase) from the {\it blind testing} phase for CLIR. In particular, the user only accesses English resources when performing the HITL interactions, and the CLIR system only accesses the English queries and user feedback, as well as the foreign language test corpus for blind testing.

The QueryBuilder system is highly effective: novice users are able to develop fine-grained queries in at most 10 minutes per query. The process is compared with a more labor-intensive, expert-driven process followed at NIST. In terms of blind CLIR performance, the generated queries from QueryBuilder result in performance that is about 12\% better than using just the overarching task for retrieval.

In summary, the primary contributions of this effort are three-fold:
\begin{itemize}
  \item The system effectively addresses the challenge of rapid and efficient corpus exploration and query generation, specifically tailored for users dealing with overarching analytic tasks
  \item A novel effective user interface (UI) which displays snippets to the user and gives the user the option to influence the system retrieval in real time based on relevance feedback. The snippets are retrieved using a combination of a fast probabilistic IR system (which uses both lexical and event-related features) and a fast neural IR system (which incorporates ``meaning'' in the retrieval).
  \item The backend (CLIR engine) of the system has been empirically tested across a diverse range of languages, demonstrating its robustness.
\end{itemize}

The paper is organized as follows: We first review related prior work in Section \ref{sec:prior_work}. We then present the overall system architecture in Section \ref{sec:querybuilder}. The underlying retrieval models (probabilistic and neural) are described in Section \ref{sec:ir_system}. The user workflow and the user interface (UI) are described in Section \ref{sec:ui-workflow}. A contrastive user workflow and UI used at NIST are described in Section \ref{sec:nist-ui-workflow}. We then present experimental results in Section \ref{sec:experiments}, and final conclusions appear in Section \ref{sec:conclusions}.

\section{Prior Work}
\label{sec:prior_work}
Our work is mostly related to {\em generating query variability} \cite{bailey2015user}, but {\em relevance feedback} \cite{salton1990improving} and {\em interactive information retrieval} \cite{tian2011active} are also related. In \cite{bailey2015user}, users are requested to re-formulate queries, according to an information need, and it is shown that the combination of these query re-formulations leads to improved performance. In \cite{salton1990improving}, it is shown that using the retrieval results of a first-pass retrieval to augment the query improves second-pass retrieval results significantly (as new terms from the top-retrieved documents, that were not present in the original queries, are now used to retrieve more relevant documents). In \cite{tian2011active}, an active-learning approach is taken, where documents that are maximally uncertain (in terms of relevance) are presented to the user for feedback. Our paper is not about {\em learning-to-rank} approaches, where user interactions help improve the ranking function \cite{liu2011learning}; we mostly focus on the process of obtaining user feedback through repeated interactions with a probabilistic IR system and a neural IR system, without having to train a ML system during those interactions (as that could slow down the process significantly).

\section{The QueryBuilder System}
\label{sec:querybuilder}
We focus on query development using an English development corpus, given an overarching task specified via a task statement. An illustration of the QueryBuilder system and its relationship to a CLIR system is presented in Figure~\ref{fig:hitl_overall}. There are two main steps in the query development workflow that we describe below.

\paragraph{Initial Query Creation:} Given an overarching task, the user uses key words or phrases (a query narrative or a list of search terms) to search an English development corpus, and identifies a small amount of representative sentences that are relevant to an aspect of the information need. By repeating this task a few times, the user comes up with an initial query (containing the search terms and a small list of example sentences). This step uses a fast, probabilistic IR system, described in Section~\ref{sec:prob-ir}.

\paragraph{Query Enrichment:} In a query-by-example setting, the user leverages a BERT-based neural IR system to retrieve sentences that are similar to the existing set of example sentences. This allows the user to quickly curate a larger and richer set of sentences, which provide more context for the query. The neural IR model is described in Section~\ref{sec:neural-ir}.

The QueryBuilder system also includes an optimized HITL user interface (UI) which allows the user to spend a very limited amount of time to perform the two steps above. We will describe the UI and the user workflow in Section~\ref{sec:ui-workflow}.

In terms of finding relevant snippets, we focus on the sentence level, which is more cost-effective than doing relevance feedback at the document level, while providing more complete information than the sub-sentence level.

\section{Information Retrieval}
\label{sec:ir_system}
We explore the complementarity of transformer-based neural IR and probabilistic IR. For the initial {\it query creation} step, we use a fast probabilistic IR model that is mostly lexical. In addition to lexical information, the probabilistic IR system~\cite{zbib2019neural} can also use neural network lexical translation and word embeddings, and can also leverage event extraction features from our Information Extraction (IE) system \cite{Boschee2005,nguyen2019one,vannguyen2022famie}. These latter features capture event trigger/agent/patient relationships detected in the query and in the corpus documents, and they lead to higher-precision retrieval. 
For the subsequent {\it query enrichment} step, we instead use a BERT-based neural IR model that works in a ``query by example" setting, thanks to its ability to capture high-level semantic features. Moreover, it is beneficial to use Neural IR to provide a complementary view of the information need.

We describe the probabilistic IR/CLIR system and the neural IR system below.

\subsection{Probabilistic IR} \label{sec:prob-ir}
The first IR component is a fast probabilistic IR model which allows interactive exploration of the large English development corpus to find example snippets of interest. It follows the probabilistic framework of \cite{miller1999hidden, zbib2019neural} which represents each query $Q$ as a set of weighted terms $Q=\{(q_1, w_1), \ldots, (q_N, w_N)\}$. The weights $\{w_1, \ldots, w_N\}$ are computed through a weighted combination of the term frequencies obtained from the various query fields. The term frequencies change from iteration to iteration, as the user highlights snippets. The term weights also change: a ``relevant to request'' score (see Table \ref{table:relevance_levels}) increases the weight by 1, while a ``relevant to task'' score increases the weight by 0.5. Finally, a ``not relevant'' score decreases the weight by 1.

Each query $Q$ is represented as a set of weighted terms $Q=\{(q_1, w_1), \ldots, (q_N, w_N)\}$. Then, given a foreign document $D$, the relevance of the document given the query is computed using the following
\begin{eqnarray}
\lefteqn{P \big(D \text{ is } \text{Rel} \mid Q \big) \propto P \big(Q \mid D \text{ is } \text{Rel} \big)} \label{eqn:prob_model1}\\
                                                     &=& \!\!\!\!\! \prod_{q_i \in Q} \Big( \alpha P(q \mid D) + \big(1-\alpha\big) P_{LM}(q) \Big)^{w_i}\label{eqn:prob_model2}\\
                                                     &=& \!\!\!\!\! \prod_{q_i \in Q} \Big( \alpha \sum_{f \in D} \frac{P(q \mid f)}{\mid D \mid} + \big(1-\alpha\big) P_{LM}(q) \Big)^{w_i} \label{eqn:prob_model3}
\end{eqnarray}
where $P(q|f)$ is the probability of translating foreign term $f$ into query term $q$;  $P_{LM}$ is a general English language model used for smoothing; and $\alpha \in (0,1]$ is an interpolation constant.

The weights $\{w_1, \ldots, w_N\}$ are computed through a weighted combination of the term frequencies obtained from the various query fields. For example, if a word $v$ has a count of $c_s(v)$ in query field $s$, and query field $s$ is assigned a weight $\theta_s$ (tuned on a development set), the overall weight for query term $v$ is given by

\begin{equation}
w(v) = \sum_{s} c_s(v) \cdot \theta_s
\label{eq:query_term_weight}
\end{equation}

Equation (\ref{eqn:prob_model3}) is used in two different IR/CLIR modes. The first mode uses lexical terms only (single words). This is done in a first-pass retrieval over the entire document collection. The second mode uses lexical terms as well as richer features that correspond to {\em predicate-argument structure} (e.g., verb and noun predicates with their thematic role arguments), detected a system such as the one in \cite{Boschee2005}. Because of the increased complexity, this is done in a second-pass retrieval over the top-retrieved documents of the first pass. Each IR/CLIR pass uses different query field weights, which are tuned on a development set.

Clearly, what are the query fields and how the corresponding weights are tuned depends on what information is available. It is very plausible for a finer-grained query  to be more accurately specified using more fields (with the overarching task being one such field) than a coarser-grained query.

The query fields available in each one of the situations we study in this paper (namely, using just the overarching task and using the user annotation through the UI) are described in the next section.


\begin{table}[t]
  \caption{Various levels of relevance given by the user.}
  \label{table:relevance_levels}
    \centering
    \begin{small}
        \begin{tabular}{p{1.6cm}|p{5.3cm}}
        \hline
Choice & Description \\
\hline
Relevant to request & Relevant to the current query (hence also relevant to the task)\\
\hline
Relevant to task & Relevant to the task but not to the query (request) \\
\hline
Neutral & The user does not want to provide a label, and does not want to see this sentence again in the next iteration \\
\hline
Not relevant & Not relevant to the task nor the request \\
        \hline
        \end{tabular}
    \end{small}
\end{table}

\setlength{\belowcaptionskip}{-5pt}
\setlength{\intextsep}{10pt plus 2pt minus 2pt}

\begin{figure}[t]
\begin{small}
{\bf task-narrative:} {\em The Flint water crisis is a public health crisis that started in 2014, after the drinking water source for the city of Flint, Michigan was changed [...]
As a result, the Flint residents were exposed to elevated levels of lead.}\\
{\bf request-narrative:} {\em Identify the events in 2014 leading to increased lead in Flint's drinking water.}\\
{\bf request-sample-doc-excerpt:} {\em The state of Michigan ran Flint’s day-to-day operations through an emergency manager, who prioritized balancing the city’s budget through a cost-cutting measure: switching Flint’s water source ...}
\end{small}
\caption{An example query.}
\label{fig:query_example}
\vspace{-0.10in}
\end{figure}

\subsection{Neural IR}
\label{sec:neural-ir}

\begin{figure*}
  \centering
  \includegraphics[scale=0.8]{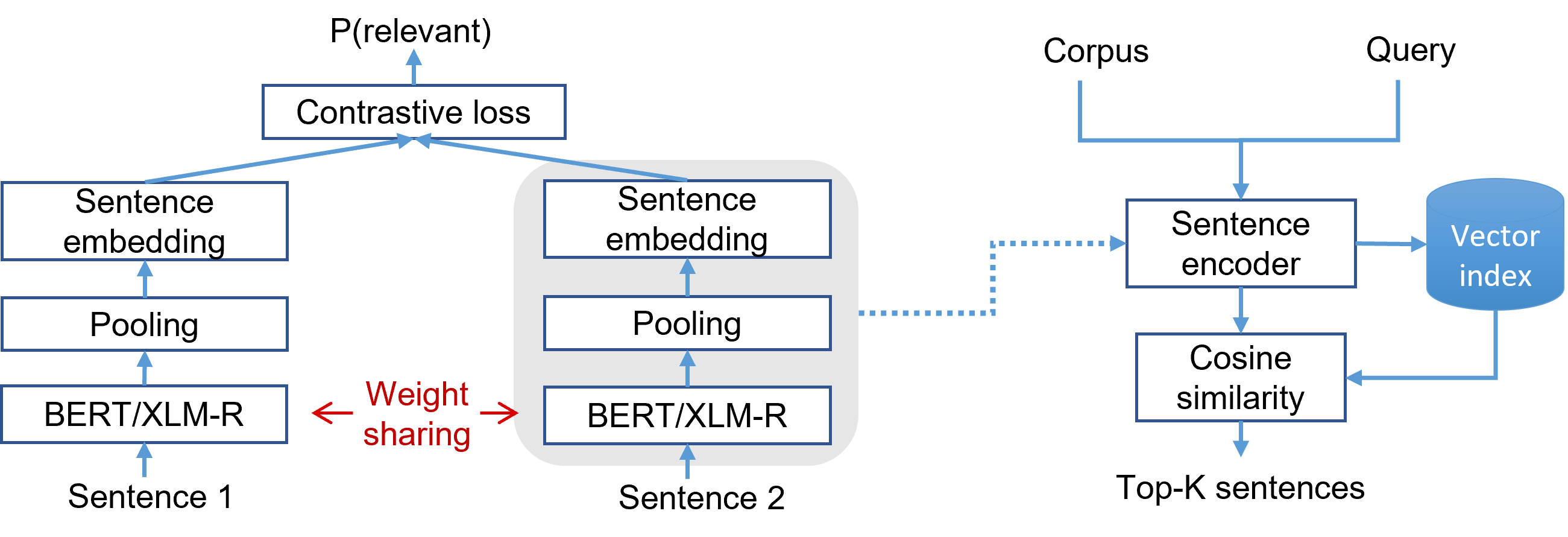}
  \caption{The Siamese Network based neural IR model. The left figure shows its internal architecture. The right-hand side shows how the trained sentence encoder can be used efficiently for retrieval.}
  \label{fig:neural_ir_model}
\end{figure*}

We develop a Siamese network based neural IR model, as shown in Figure~\ref{fig:neural_ir_model} (left). This model takes a pair of sentences as input and uses a pre-trained BERT \cite{devlin-etal-2019-bert} or XLM-R \cite{conneau-etal-2020-unsupervised} model to encode each sentence. BERT is a powerful language model that has achieved state-of-the-art results on various NLP tasks. XLM-R extends BERT with cross-lingual support, making it suitable for multilingual IR tasks. After encoding the sentences, the model extracts a pair of sentence embeddings by pooling the outputs of the encoder model. It then predicts whether the sentences in the pair are relevant or not.

This model is trained with positive (relevant) or negative (irrelevant) pairs with contrastive loss. At inference time, since the encoder for query sentences and passages is the same, we  simply calculate the cosine similarity of a pair of sentence embeddings. Therefore, this approach is very efficient: we only need to calculate the embeddings of query or passages once, and then retrieval is fast, done in real time using an efficient vector index. Figure~\ref{fig:neural_ir_model} (right) shows the inference process, given a corpus and a query.


We use neural IR models trained with MS MARCO and Paraphrase datasets, respectively. The MS MARCO model is shown to be useful for few-shot IR, while the paraphrase model allows us to explore alternative expressions of the information need, which provide more lexical and syntactic diversity. For the QueryBuilder system, we chose the paraphrase model because it performed significantly better on our development queries. We use the pretrained paraphrase model~\footnote{\url{https://huggingface.co/sentence-transformers/paraphrase-distilroberta-base-v1}} from Huggingface, which is trained from millions of examples for paraphrases, mainly derived from Wikipedia, from news reports, AllNLI-entailment pairs, etc.


\section{The User Interface and Workflow}
\label{sec:ui-workflow}
The user is first given the task description, in the form of a ``narrative'', to understand the overarching task. In an operational scenario, given the overarching task, the user would form his/her own sub-topics, based on which QueryBuilder would help with the actual query formation. Then, to evaluate the query building process, we would have to collect relevance judgments from the top-retrieved documents per query. Since we did not have the resources to perform such judgments, we ``borrowed'' the sub-topics from the expert NIST annotators. To do that, our users were allowed to look at the expert-provided ``narrative'' description of each sub-topic; we then proceeded as if they had come up with the sub-topics themselves. To make the task as realistic as possible, the users were not allowed to use these sub-topic descriptions verbatim when searching the English collection; they had to paraphrase the descriptions and use their own wording, as if {\em they came up with the sub-topics themselves}. Figure \ref{fig:query_example} shows an example of a task narrative, request (sub-topic) narrative, and excerpt highlighted by an experienced annotator.

Note that, in the downstream CLIR experiments that used the output of QueryBuilder, we did not use any of the "narrative" fields provided by the expert annotators. The CLIR experiments only used narratives that our own users came up with, as well as sentences extracted from automatically retrieved documents that used the manual narratives as queries.


{\bf Initial query creation}. The user first enters a search phrase or a list of search terms that he believes are relevant to the sub-topic of interest and presses a button to run the probabilistic IR system, which retrieves sentences from the English development corpus. The user then inspects the top-ranked retrieved sentences for relevance\footnote{The user has the option to try the search multiple times, with different search terms, if the top-retrieved results do not look relevant. The QueryBuilder system records which queries lead to good results, as indicated by the user.}. The user finally selects typically $\leq 5$ sentences that seem relevant to the sub-topic. This step results in an initial query consisting of: (i) the search terms entered by the user, and (ii) a handful of sentences that seem relevant to the sub-topic of interest (request). This process usually takes 2-4 minutes per query.

{\bf Query enrichment}. The goal is to find more sentences that are relevant to the sub-topic. The user clicks a button to run neural IR, and retrieves a ranked list of sentences that are similar to the previously selected sentences (from the first step). The user can do this iteratively - the later iterations are expected to provide better results because they will utilize all previously selected sentences as the ``query''. The user then inspects top-10 retrieved sentences in each iteration to find up to 25 sentences that seem relevant to the sub-topic. This process usually takes less than 7 minutes per query.




\section{Query Generation at NIST and Proposed Improvements}
\label{sec:nist-ui-workflow}
To answer the question ``how does the system compare to existing systems?'' we briefly describe the NIST workflow for generating analytic tasks and subtopic queries for used in the IARPA BETTER IR datasets \cite{soboroff-etal-2023-better}. In order to be useful evaluation queries, they need to pertain to a topic for which relevant information appears in the target collection, but making sure that the number of relevant documents is not too high (queries with too many relevant documents do not allow for a useful comparison between systems \cite{buckley2007bias}).
As such, the process at NIST is designed to scope out useful search topics, providing a sense of how prevalent it is in the broader collection without exhaustive searching.


The assessors are provided a search tool based on ElasticSearch.\footnote{https://elastic.co}  This tool supports searches over both English and foreign-language documents.  In the English collection, the documents have been automatically annotated for entities using Spacy\footnote{https://spacy.io}, and search results are faceted among people, places, organizations, and geopolitical entities.  In both languages, the search interface allows simple operators including minus to indicate negation and quotes to indicate phrases.

To support the need to collect sample relevant documents with snippets identified, the English side of the system allows the user to highlight a passage of text per document.  When a passage is highlighted, the document and passage are recorded as relevant.  A ``topic editor'' tab allows the user to define the scope of their search interest, and relevant documents are collected either as part of an overall task or an individual query, depending on the user's activity in the topic editor.

The NIST query development proceeds as follows. (1) The user explores the English collection in order to identify a usable overall analytic task. (2) Next, the user defines 5-10 individual search requests that contribute to their goal of understanding the overall analytic task. For example, for the Flint water crisis task shown in Figure \ref{fig:query_example}, these queries might concern what decisions led to the water contamination, how many people were harmed, etc.
(3) For each query, they perform a foreign-language corpus search to determine that less than 15 of the top 20 documents retrieved for that search are relevant to the query, revising their query as needed to meet the appropriate scope. (4) They compose a sentence-length search request corresponding to the query. (5) They find two relevant English documents that contain answers to the query, which they highlight.

It takes around an hour to complete an analytic task. As can be seen, this process involves a significant amount of trial and error: the user has to manually enter search terms (frequently multiple times) until he retrieves enough relevant documents. The process could be significantly improved by adding the QueryBuilder automation described earlier:

\begin{enumerate}[labelindent=0pt,labelwidth=!,leftmargin=*]
\setlength{\itemsep}{0pt}
\setlength{\parskip}{0pt}
\item Highlights could be used in subsequent searches to guide the system toward more accurate/relevant retrieval.
\item Allowing users to use a "relevance score" in the judgment (instead of just highlighting snippets as relevant), for finer-grained re-weighting of terms in the IR engine.
\item Incorporation of query-by-example capability, which takes ``meaning'' into account, based on NN-based embeddings.
\end{enumerate}


\section{Experiments}
\label{sec:experiments}
We first evaluated our system on an Arabic-English CLIR task, using queries and documents that were made available by NIST for the IARPA BETTER Phase-1 IR datasets .
There are eight overarching analytic tasks, each one consisting of a variable number of sub-topics (analytic requests), ranging between five and nine. In total, there are 54 analytic requests. The English document collection consists of about 750K news articles. Two human subjects (users), without significant experience in annotation tasks or in IR, participated in this study.

As mentioned earlier, at the beginning of the process, the users are provided with the task title and narrative, to get an idea of the task. Then, for each one of the analytic requests (sub-topics of task), they read the corresponding NIST-provided narrative of the request, but they are instructed not to memorize it. Instead, they re-formulate it and come up with their own search terms with which they search the English corpus.

%
%

Various statistics from the query generation and the query enrichment stages are shown in Table \ref{table:stats_initial_query_creation} and Table \ref{table:stats_query_enrichment} of Appendix \ref{sec:appendix_stats}, respectively, along with high-level conclusions.

\begin{table}[ht]
\caption{nDCG results using queries from the 1st annotation stage (initial query creation), the 2nd annotation stage (query enrichment), and from NIST.}
\label{table:ir_results}
\begin{center}
\begin{tabular}{|r|c|c|c|} \hline
					&						& {\bf query}	& {\bf nDCG}	\\ \hline \hline
\multirow{ 3}{*}{\bf user 1}	& \multirow{ 2}{*}{1st stage}	& search terms	& 0.356	\\ \cline{3-4}
					&						& + sel. sent.	& 0.536	\\ \cline{2-4}
					& 2nd stage				& + sel. sent.	& 0.546	\\ \hline \hline
\multirow{ 3}{*}{\bf user 2}	& \multirow{ 2}{*}{1st stage}	& search terms	& 0.478	\\ \cline{3-4}
					&						& + sel. sent.	& 0.533	\\ \cline{2-4}
					& 2nd stage				& + sel. sent.	& 0.540		\\ \hline \hline
\multirow{ 2}{*}{\bf NIST}	& \multicolumn{2}{|c|}{just overarching task}	& 0.426	\\ \cline{2-4}
					& \multicolumn{2}{|c|}{+ req. narrative+excerpt}	& 0.628		\\ \hline
\end{tabular}

\end{center}
\end{table}

Normalized discounted cumulative gain (nDCG) results on an Arabic-English CLIR task using the probabilistic IR system are shown in Table \ref{table:ir_results}. The document collection consists of about 865K Arabic documents and the probabilistic IR system uses the queries generated in the two stages mentioned above. In the 1st stage, two results are shown: using only the search terms entered by the users, and additionally using the marked sentences. As can be seen, results with the queries generated from selected sentences are substantially better (6-18\% absolute) than using just the search terms during the first stage. This shows that the process of sentence retrieval, annotation (relevance feedback) and term re-weighting based on that feedback works well in QueryBuilder and gives gains. Also, results from the second stage (query enrichment) are a little better (about 1\% absolute) than results from the first stage (initial query creation). For comparison, results obtained with just the overarching task as the query (so, no sub-topic distinction) and results with the queries generated by expert annotators from NIST (as described in Section \ref{sec:nist-ui-workflow}) are shown at the bottom of the table. In almost all cases, the queries generated by our novice annotators improved upon the task-only queries, with the maximum improvement being almost 12\% absolute, while the difference from the expert annotators is about 8\% absolute.

\subsection{Query Combination}
Additionally, we investigated methods for improving the performance of the CLIR system by combining existing queries with those generated using QueryBuilder. This query combination approach aims to leverage the strengths of both query sources to improve retrieval performance. We used the data developed for the IARPA BETTER program encompassing various languages and diverse task structures. In addition to data described in the Arabic-English CLIR task, for the Farsi-English task, there are 10 overarching analytic tasks with 53 analytic requests with over 850K retrieval corpus and for Russian, Chinese, and Korean there are 10 overarching analytic tasks with 48 analytic requests and 1M retrieval corpus. Full dataset details are provided in \cite{soboroff-etal-2023-better}(available at \url{https://ir.nist.gov/better/}).
 The results of this query combination are shown in Table \ref{tab:hitl_results}  which containes nDCG scores on five different languages: Arabic, Farsi, Chinese, Korean, and Russian. The performance of the CLIR system is shown for three different query configurations: (i) Using only sample documents selected by experts; (ii) Similar to (i) but additionally including a detailed description of the information need; (iii) Combining the queries of (ii) with those generated by QueryBuilder.

 \begin{table}[]
  \caption{CLIR System Performance with Different Query Configurations and multiple languages: Arabic (ar), Farsi (fa), Chinese (zh), Korean (ko), and Russian (ru)}
  \begin{tabular}{|l|c|c|c|c|c|}
  \hline
  \multicolumn{1}{|c|}{} & \textbf{ar} & \textbf{fa} & \textbf{zh} & \textbf{ko} & \textbf{ru} \\ \hline
  Sample Documents       & 0.54        & 0.646       & 0.324       & 0.585       & 0.389       \\ \hline
  + Detailed Description & 0.582       & 0.646       & 0.342       & 0.577       & 0.404       \\ \hline
  + QueryBuilder         & 0.600         & 0.643       & 0.355       & 0.595       & 0.415       \\ \hline
  \end{tabular}

  \label{tab:hitl_results}
  \end{table}
As expected, the more information about the search task is made available to the system the more performance improves. As mentioned earlier, the user interacted with QueryBuilder for only 7 minutes per search task. The results demonstrate the effectiveness of QueryBuilder at providing the user with an efficient, interactive way of further improving the quality of CLIR system.

%





\section{Concluding Remarks}
\label{sec:conclusions}
This paper presented a novel, interactive system called {\em QueryBuilder}, which allows a novice, English-speaking user to create queries with a small amount of effort, via exploration of an English development corpus. The QueryBuilder system uses efficient IR algorithms (probabilistic and BERT-based) and retrieves sentence-level information, which the user can easily access and annotate.
Our experiments showed that with a small amount of effort (at most 10 minutes per sub-topic), novice users are able to form good fine-grained queries. CLIR results from an Arabic-English retrieval task show that the performance of the queries created by the novice users is not too far behind that of more experienced annotators. Additionally, we explored methods for query combination, leveraging both expert-curated and QueryBuilder-generated queries to improve CLIR performance. The results underscored the system's potential for enhancing information retrieval across various languages. Overall, QueryBuilder presents a promising approach to streamline the query development process, offering potential benefits to analysts and researchers seeking efficient and effective information retrieval.

\section*{Acknowledgments}
This research is based upon work supported in part by the Office of the Director of National Intelligence (ODNI), Intelligence Advanced Research Projects Activity (IARPA), via 2019-19051600006 under the Better Extraction from Text
Towards Enhanced Retrieval (BETTER) Program. We would like to thank members of the Analytics and Machine Intelligence (AMI) department at Raytheon BBN for useful discussions and support. Special thanks to  Brian Ulicny for his valuable feedback and suggestions and to Sean Colbath from AMI for his help with creating the demo video.

\section*{Disclaimers}
The views and conclusions contained herein are those of the authors and should not
be interpreted as necessarily representing the official policies, either expressed or implied, of ODNI, IARPA,
or the U.S. Government. The U.S. Government is authorized to reproduce and distribute reprints for
governmental purposes notwithstanding any copyright annotation therein.
This paper may mention certain companies, products, or services.  Any such mention does not imply a recommendation or endorsement for that company, product, or service by NIST.

\bibliography{bibliography}
\bibliographystyle{ACM-Reference-Format}
  \begin{table*}[!h]
\caption{Various statistics calculated from the 2nd annotation stage (query enrichment).}
\label{table:stats_query_enrichment}
\begin{center}
\begin{tabular}{|r|l|c|c|c|c|c|c|} \hline
					&				& iter. 1	& iter. 2	& iter. 3	& iter. 4	& iter. 5	& iter. 6+	\\ \hline
\multirow{ 2}{*}{user 1}	& (i) \ \# ``rel.'' sent.		& 3.41			& 5.63			& 15.22			& 19.17			& 10.66			& 13.0			\\ \cline{2-8}
					& (ii) ``rel.'' sent. len.	& 31.72			& 32.74			& 33.31			& 32.25			& 38.06			& 40.46			\\ \hline
\multirow{ 2}{*}{user 2}	& (i) \ \# ``rel.'' sent.		& 3.10			& 5.00			& 7.64			& 4.22			& 2.88			& 2.61			\\ \cline{2-8}
					& (ii) ``rel.'' sent. len.	& 32.97			& 32.55			& 32.93			& 34.24			& 32.80			& 33.13			\\ \hline
\end{tabular}

\end{center}
\end{table*}

\appendix

\section{Statistics from the Query Generation Process with QueryBuilder}
\label{sec:appendix_stats}
Table \ref{table:stats_initial_query_creation} shows various statistics from the first stage of query generation with QueryBuilder. Specifically, it shows (i) for each search iteration, the average number of words that the users enter into the search box. (ii) The number of analytic requests for which they run exactly one or two iterations. (iii) The average number of sentences marked as ``relevant'', per iteration. (iv) The average length of the sentences marked as relevant. High-level observations from these results are: (a) it is enough to enter search keywords only once for the majority of the requests (74\% of the requests for user 1 and 63\% for user 2). (b) Both users mark about 3-4 sentences as ``relevant'' on average, although user 2 performs search with more keyword terms (8-9 vs 4 of user 1).
Various statistics from the second stage are shown in Table \ref{table:stats_query_enrichment}. Specifically, (i) The average number of sentences marked as ``relevant''. (ii) The average length of the sentences marked as relevant. Some high-level observations are: (i) Interestingly, the number of sentences marked as ``relevant'' seems to reach a peak at around 3-4 iterations for both users. (ii) The average sentence length of the marked sentences varies between 32 and 40 for user 1 while it remains pretty constant at 32-33 for user 2.

\begin{table}[h]
  \caption{Various statistics calculated from the 1st annotation stage (initial query creation).}
  \label{table:stats_initial_query_creation}
  \begin{center}
  \begin{tabular}{|r|l|c|c|} \hline
            &				& {\bf iter. 1}	& {\bf iter. 2}	\\ \hline
  \multirow{ 4}{*}{\bf user 1}	& (i) \ \# search terms	& 4.07			& 4	\\ \cline{2-4}
            & (ii) \ \# requests		& 40			& 11	\\ \cline{2-4}
            & (iii) \ \# ``rel.'' sent.		& 3.92			& 4.25	\\ \cline{2-4}
            & (iv) ``rel.'' sent. length 	& 31.43			& 33.57	\\ \hline
  \multirow{ 4}{*}{\bf user 2}	& (i) \ \# search terms		& 8.43			& 9	\\ \cline{2-4}
            & (ii) \ \# requests			& 34			& 19	\\ \cline{2-4}
            & (iii) \ \# ``rel.'' sent.			& 3.09			& 3.68	\\ \cline{2-4}
            & (iv) ``rel.'' sent. length		& 31.54			& 34.07	\\ \hline
  \end{tabular}
  
  \end{center}
  \end{table}

From the above statistics we can conclude that the amount of work the QueryBuilder users have to do in terms of reading and annotating passages is not high, and clearly much less than reading entire documents.

\section{Screenshots from QueryBuilder}
\label{sec:appendix_screenshots}
\begin{figure*}
  \centering
  \includegraphics[scale=0.30]{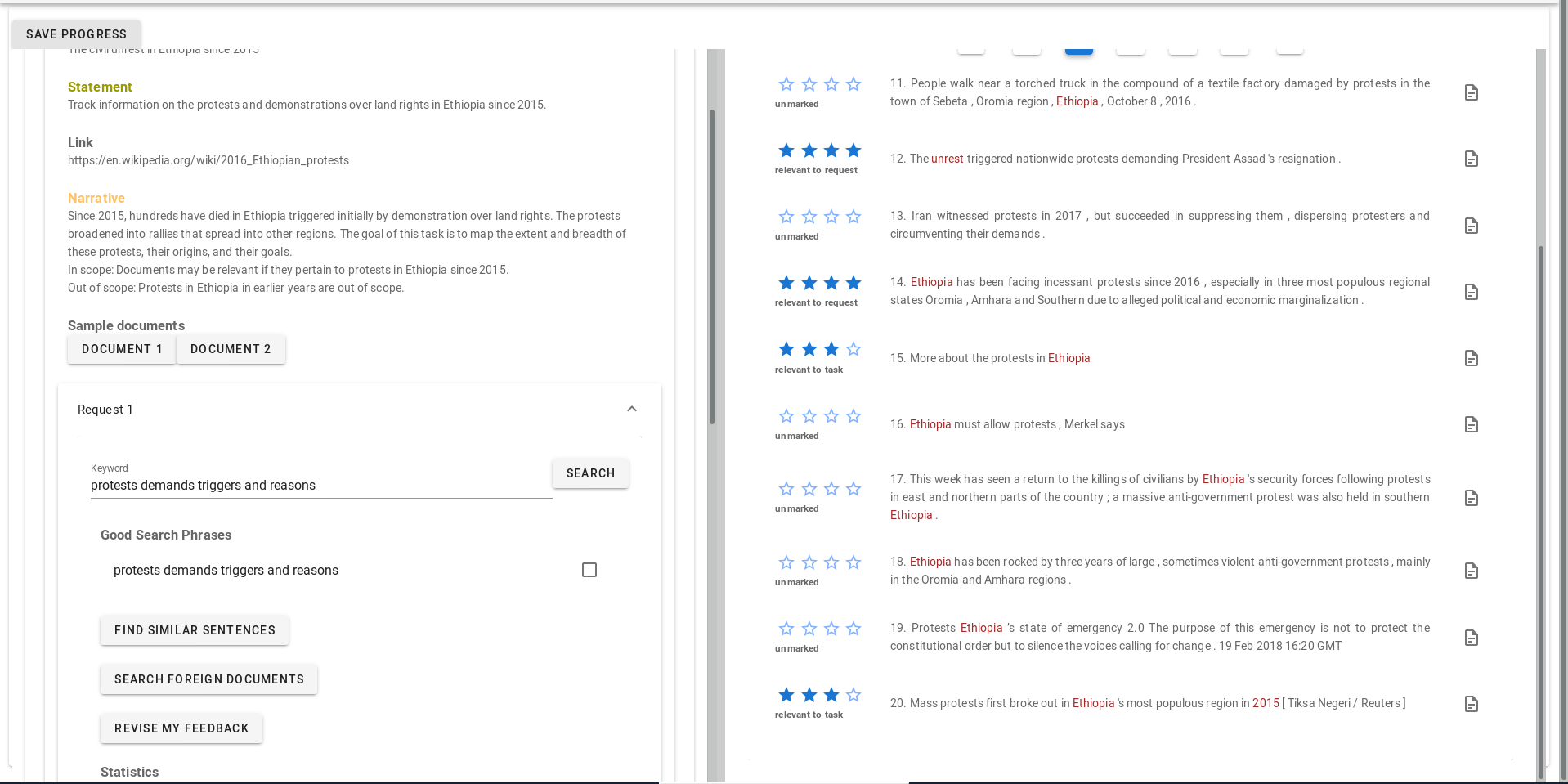}\\(a)\\[0.2in]
  \includegraphics[scale=0.30]{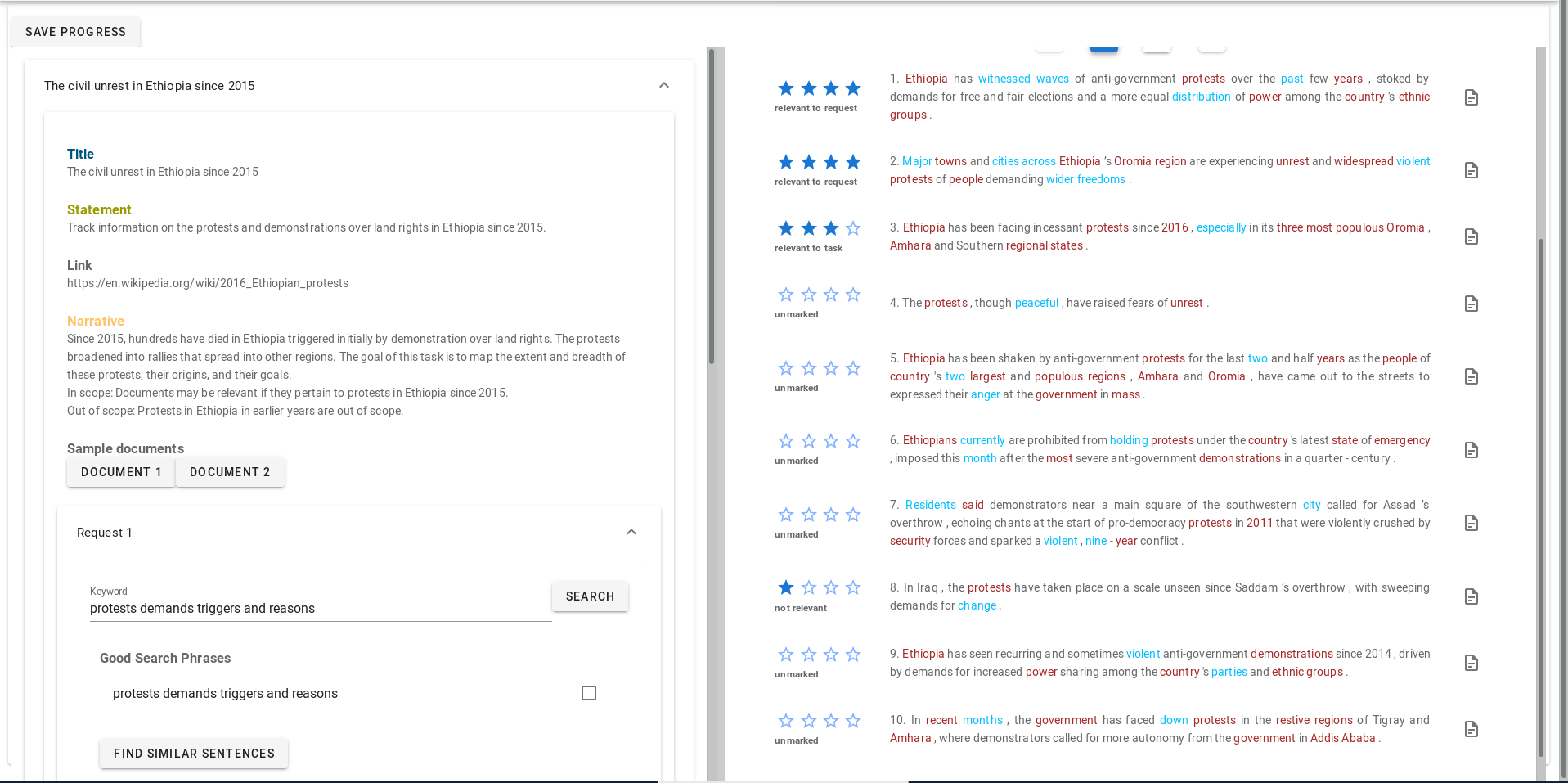}\\(b)\\
  \caption{Screenshots from stages 1 and 2 of QueryBuilder. Screenshot (a) shows the search terms that the user entered into the search box on the left-hand side and the retrieved sentences returned by the probabilistic system on the right-hand side. The words in red are query words (after query expansion). The relevance score (with the stars) given by the user is also shown for four of the retrieved sentences. Screenshot (b) shows the results obtained with the neural IR system on the right-hand side. Words in blue are query words that the system matched in addition to the previously matched words (in red).}
  \label{fig:querybuilder_screenshots}
\end{figure*}

Two screenshots from the QueryBuilder UI are shown in Figure \ref{fig:querybuilder_screenshots}. Screenshot (a) shows the search terms that the user entered into the search box on the left-hand side and the retrieved sentences returned by the probabilistic system on the right-hand side. The user has the ability to annotate each one of the returned results with one of the relevance levels shown in Table \ref{table:relevance_levels}, using the corresponding number of stars. Screenshot (b) shows the results obtained with the neural IR system (again, on the right-hand side); these are sentences that have a ``meaning'' similar to the marked sentences of the first stage (according to the learned embedding).

\end{document}